\documentclass[10pt]{article}
\usepackage[dvips]{graphicx}

\usepackage{epsfig}
\usepackage{graphicx}
\usepackage{indentfirst}
\usepackage{amsmath}
\usepackage{amsfonts}
\usepackage{amssymb}

\begin{document}

\begin{center}
\begin{flushright}\begin{small}    UFES 2012
\end{small} \end{flushright} \vspace{1.5cm}
\huge{Planar Symmetry in f(T) gravity} 
\end{center}

\begin{center}
{\small  Manuel E. Rodrigues $^{(a)(b)}$}\footnote{E-mail
address: esialg@gmail.com},
{\small    M. J. S. Houndjo $^{(c)(d)}$}\footnote{E-mail address:
sthoundjo@yahoo.fr},
{\small   D. Momeni $^{(e)}$}\footnote{E-mail address:
d.momeni@yahoo.com} 
 and
{\small    R. Myrzakulov $^{(e)}$}\footnote{E-mail address:
rmyrzakulov@csufresno.edu} 
\vskip 4mm

(a) \ Universidade Federal do Esp\'{\i}rito Santo \\
Centro de Ci\^{e}ncias
Exatas - Departamento de F\'{\i}sica\\
Av. Fernando Ferrari s/n - Campus de Goiabeiras\\ CEP29075-910 -
Vit\'{o}ria/ES, Brazil \\
(b) \ Faculdade de F\'{\i}sica, Universidade Federal do Par\'{a}, 66075-110, Bel\'em, Par\'{a}, Brazil\\

(c)\ Departamento de Engenharia e Ci\^{e}ncias Naturais - CEUNES \\
Universidade Federal do Esp\'irito Santo \\  CEP 29933-415 - S\~ao Mateus/ ES, Brazil.\\
\ (d) Institut de Math\'{e}matiques et de Sciences Physiques (IMSP)\\01 BP 613 Porto-Novo, B\'{e}nin\\
\ (e) Eurasian International Center for Theoretical Physics\\
Eurasian National University, Astana 010008, Kazakhstan\\
\vskip 2mm

\end{center}

\begin{abstract}
Exact solutions for planar Weitzenboch geometries in generalized teleparallel gravity with torsion is presented. The model predicted the existence of Moller, Kattler-Wittaker and planar de-Sitter geometries for different forms of action.

\end{abstract}
Pacs numbers: 04.50. Kd, 04.70.Bw, 11.30.-j

\section{Introduction}
Planar symmetry means the spacetime has at least two non commutative horizon generators $\zeta_x,\zeta_y$. So by fixing this symmetric structure, the static configurations in time independent basis are functions of a single axial coordinate $z$. In general relativity the solution for field equations in the case of a  hypersurface orthogonal but homogeneous spacetime reduces to an integrable system of differential equations which has been investigated previously (for a complete list of all classical of solutions, see Sec.15 of \cite{exactsolutions}). From string theory we also need such plane gravitating objects as the source of the plane domain walls. The geometry of the plane has more freedom for consideration in comparison to the localized sources with associated symmetry of sources. The axial symmetry on a non compact sub-space reduces to the plane symmetry. In modified gravity the plane symmetry also plays an important role as the preferred symmetry in some cases. In  modification of the usual Einstein gravity by curvature terms, the so called $f(R)$ gravity, proposed recently (see \cite{review} and references in it) the planar solutions have been investigated \cite{sharif}. In non-relativistic regime of gravity, for example in Horava-Lifshitz gravity \cite{Setare:2010gy}-\cite{Momeni:2009au}, this type of solution has been proposed to solve the problem of negative cosmological constant. Also, another kinds of planar solutions in relation to the monopole in general relativity have been investigated \cite{nouri}.
Recently, the generalized teleparallel gravity achieves great success in the description of the gravity, not as a effect of curvature but as the effect of torsion \cite{fT}-\cite{BuenoSanchez:2010wd}. Different kinds of exact solutions has been investigated in the framework of this new alternative gravity . In this paper we want to derive the field equations for planar symmetry in f(T) gravity and investigate some possible solutions which have more physical meanings than others known until now. So, our point of view is from the physical importance, not pure mathematical aspects.
The plan of this work is as follows:
In section II, we review the static perfect fluids in general relativity (GR). In sections III-IV the formalism of f(T) generalized teleparallel  is motivated for plane symmetry. In section V, the reduction formalism is presented and contains a significant fact about the non equivalence of teleparallel TEGR and GR at level of action in plane symmetry. In section VI, the vacuum solutions in absence of an effective cosmological constant is obtained. Also, we proved the existence of Moller space time in our model. The section VII is devoted to the study the constant torsion solutions motivated by the existence of a planar de-Sitter solution. Section VIII presented the Kottler family of solutions as a singularity free solution. The conclusion is presented in the last section.

\section{Static Plane-symmetric perfect fluid solutions in GR}
The general form of a static plane symmetric solution is defined by \cite{Taub},
\begin{eqnarray}
ds^2=-e^{2\nu}dt^2+z^{2}(dx^2+dy^2)+zF(z)^{-1}dz^2\,\,,
\end{eqnarray}
where it has been assumed that this metric has a perfect fluid source with isentropical components with the fundamental equation of state $p=p(\rho)$ or $\rho=\rho(p)$ where here $p$ is the principal pressure component with the maximum eigenvalue of the energy momentum tensor along a timelike hypersurface orthogonal spacetime and $\rho$ the energy component which satisfies the null energy condition as well. The Einstein field equations give us the following two solvable differential equations \cite{exactsolutions},
\begin{eqnarray}
\frac{2zp'}{\rho(p)+p}&=&1-\kappa_{0}p\frac{z^3}{F}=-2z\nu' \,\,,\\
F'&=&-\kappa_{0}\rho(p)z^2\,\,.
\end{eqnarray}
If we know the equations of state then integration straightforwardly gives us the metric functions $F(z),\nu(z)$. Several kinds of the reconstruction mechanism can be used here to find new solutions. Also, the system exhibits a conservation equation in terms of the pressure and the reduced metric function as the following
\begin{eqnarray}\nonumber
\frac{p'}{p}&=&\frac{z^2+G'}{G}\frac{G+z^3}{z^3-G}\,\,,\nonumber\\
G&=&-\frac{F(z)}{\kappa_{0} p(z)}\,\,.
\end{eqnarray}
So, if we  prescribe the explicit form of $G(z)$ from the continuity equation we can obtain $p(z)$ and $\nu=\nu(z)$ as quadrators. Also the reconstruction of the exact solutions using the equation of state is possible. The most simplest case is a barotropic one with $p = (\gamma-1)\rho$. Specially, for  $\rho = const$ the solutions have been obtained \cite{Taub}. Such static solutions with plane-symmetric perfect fluids can be investigated as special algebraic  sub-cases of the Levi-Civita static cylindrically-symmetric solutions.

\section{\large The basic concepts from $f(T)$ theory}
For a general spacetimes metric, we can define the line element as
\begin{equation}
dS^{2}=g_{\mu\nu}dx^{\mu}dx^{\nu}\; .
\end{equation} 
 This metric can be projected in the tangent space to the manifold, using the representation of the tetrad matrix, where the line element is, 
\begin{eqnarray}
dS^{2} &=&g_{\mu\nu}dx^{\mu}dx^{\nu}=\eta_{ij}\theta^{i}\theta^{j}\label{1}\; ,\\
dx^{\mu}& =&e_{i}^{\;\;\mu}\theta^{i}\; , \; \theta^{i}=e^{i}_{\;\;\mu}dx^{\mu}\label{2}\; ,
\end{eqnarray} 
where $\eta_{ij}=diag[1,-1,-1,-1]$ and $e_{i}^{\;\;\mu}e^{i}_{\;\;\nu}=\delta^{\mu}_{\nu}$ or  $e_{i}^{\;\;\mu}e^{j}_{\;\;\mu}=\delta^{j}_{i}$. The square root of the metric determinant is given by  $\sqrt{-g}=\det{\left[e^{i}_{\;\;\mu}\right]}=e$.  Now, we describe the spacetime through the tetrad matrix and then define the Weitzenbock's connection as \cite{hayashi}
\begin{eqnarray}
\Gamma^{\alpha}_{\mu\nu}=e_{i}^{\;\;\alpha}\partial_{\nu}e^{i}_{\;\;\mu}=-e^{i}_{\;\;\mu}\partial_{\nu}e_{i}^{\;\;\alpha}\label{co}\; .
\end{eqnarray}

\begin{eqnarray}
T^{\alpha}_{\;\;\mu\nu}&=&\Gamma^{\alpha}_{\nu\mu}-\Gamma^{\alpha}_{\mu\nu}=e_{i}^{\;\;\alpha}\left(\partial_{\mu} e^{i}_{\;\;\nu}-\partial_{\nu} e^{i}_{\;\;\mu}\right)\label{tor}\;.
\end{eqnarray}

\begin{eqnarray}
K^{\mu\nu}_{\;\;\;\;\alpha}&=&-\frac{1}{2}\left(T^{\mu\nu}_{\;\;\;\;\alpha}-T^{\nu\mu}_{\;\;\;\;\alpha}-T_{\alpha}^{\;\;\mu\nu}\right)\label{cont}\; ,\\
S_{\alpha}^{\;\;\mu\nu}&=&\frac{1}{2}\left( K_{\;\;\;\;\alpha}^{\mu\nu}+\delta^{\mu}_{\alpha}T^{\beta\nu}_{\;\;\;\;\beta}-\delta^{\nu}_{\alpha}T^{\beta\mu}_{\;\;\;\;\beta}\right)\label{s}\;.
\end{eqnarray}
Through  (\ref{tor})-(\ref{s}), we define the scalar torsion scalar as  
\begin{eqnarray}
T=T^{\alpha}_{\;\;\mu\nu}S^{\;\;\mu\nu}_{\alpha}\label{tore}\; .
\end{eqnarray}

\begin{eqnarray}
S[e^{i}_{\mu},\Phi_{A}]=\int\; d^{4}x\;e\left[\frac{1}{16\pi}f(T)+\mathcal{L}_{Matter}\left(\Phi_{A}\right)\right]\label{action}\; ,
\end{eqnarray}
where we used the units $G=c=1$ and the $\Phi_{A}$ are the matter fields. Considering the action (\ref{action}) as a functional of the fields  $e^{i}_{\mu}$ and $\Phi_{A}$,  and vanishing the variation of the functional  with respect to the field  $e^{i}_{\nu}$, i.e. the principle of minimum action, one obtains the following equation of motion \cite{Bengochea:2008gz}
\begin{eqnarray}
S^{\;\;\nu\rho}_{\mu}\partial_{\rho}Tf_{TT}+\left[e^{-1}e^{i}_{\mu}\partial_{\rho}\left(ee^{\;\;\alpha}_{i}S^{\;\;\nu\rho}_{\alpha}\right)+T^{\alpha}_{\;\;\lambda\mu}S^{\;\;\nu\lambda}_{\alpha}\right]f_{T}+\frac{1}{4}\delta^{\nu}_{\mu}f=4\pi\mathcal{T}^{\nu}_{\mu}\label{em}\; ,
\end{eqnarray}
where $\mathcal{T}^{\nu}_{\mu}$ is the energy momentum tensor, $f_{T}=d f(T)/d T$ and $f_{TT}=d^{2} f(T)/dT^{2}$. If we consider $f(T)=a_{1}T+a_{0}$, the TT is recovered with a cosmological constant.
\par
In the next section, we will make some considerations for the manifold symmetries in order to obtain simplifications in the equations of motion and the specific solutions of these symmetries.

\section{\large   Planar geometry}

We consider from the beginning the tetrads matrix as the fundamental fields of the $f(T)$ theory. Now, in the same way that the frames were constructed for the TT theory in \cite{maluf1} and \cite{maluf2,maluf3}, our tetrad ansatz is elaborated by fixing the degree of freedom as follows  $e_{0}^{\;\;\mu}=u^{\mu}$, where $u^{\mu}$ is the four velocity of an observer in free fall, and $e_{1}^{\;\;\mu},e_{2}^{\;\;\mu}$ and $e_{3}^{\;\;\mu}$ are oriented along the unitary  vectors in the Cartesian directions $x$ , $y$ and $z$. For the planar symmetry,  as seen in \cite{cai} the result is given by the diagonal matrix
\begin{eqnarray}
\left\{e^{i}_{\;\;\mu}\right\}= \left[ \sqrt{A(r)},\sqrt{B(r)},\sqrt{C(r)},\sqrt{C(r)}\right]\label{tetra}\; ,
\end{eqnarray}

Using the relations (\ref{1}) and (\ref{2}), one can write the components of the metric, through the line element of planar and static spacetimes as
\begin{equation}
dS^{2}=A(r)dt^{2}-B(r)dr^{2}-C(r)\left(dx^2+dy^2\right)\label{ele}\; ,
\end{equation} 
with $r=|z|$ and $z$ is the Cartesian coordinate. This choice of tetrad matrices is not unique, because the aim of letting the line element invariant under local Lorentz transformations is to obtain the form (\ref{1}). Using  (\ref{tetra}), one can obtain $e=\det{\left[e^{i}_{\;\;\mu}\right]}=\sqrt{AB}C$.
The components of the torsion (\ref{tor}), contorsion (\ref{cont}) and tensor $S_{\alpha}^{\mu\nu}$ (\ref{s}) are give by
\begin{eqnarray}
T^{0}_{\;\;10}&=&\frac{A^{\prime}}{2A}\,,\,T^{2}_{\;\;12}=T^{3}_{\;\;13}=\frac{C^{\prime}}{2C}\,,\,K^{01}_{\;\;\;\;0}=\frac{A^{\prime}}{2AB}\,,\,K^{21}_{\;\;\;\;2}=K^{31}_{\;\;\;\;3}=\frac{C^{\prime}}{2BC}\,,\\
S_{0}^{\;\;10}&=&\frac{C^{\prime}}{2BC}\,,\,S_{2}^{\;\;12}=S_{3}^{\;\;13}=\frac{C^{\prime}}{4BC}+\frac{A^{\prime}}{4AB}\,.
\end{eqnarray} 
where the prime  ($^{\prime}$) denotes the derivative with respect to  the coordinate $r$. The torsion scalar (\ref{tore}) is
\begin{eqnarray}
T=\frac{\left(C^{\prime}\right)^2}{2BC^2}+\frac{A^{\prime}C^{\prime}}{ABC}\;\label{T}.
\end{eqnarray}

The energy-momentum tensor is
\begin{equation}
\mathcal{T}^{\mu}_{\nu}=diag\left[\rho(r),-p_{r}(r),-p_{x}(r),-p_{y}(r)\right]\;.
\end{equation}
One can now re-write the equations of motion (\ref{em}) for the planar symmetry as 
\begin{eqnarray}
4\pi\rho &=& \frac{f}{4}-\left(\frac{C^{\prime\prime}}{2BC}+\frac{A^{\prime}C^{\prime}}{4ABC} -\frac{B^{\prime}C^{\prime}}{4B^2C}\right)f_T-\frac{C^{\prime}}{2BC}\left(f_T\right)^{\prime}\,,\label{dens} \\
4\pi p_{r} &=&  \left[\frac{\left(C^{\prime}\right)^2}{4BC^2}+\frac{A^{\prime}C^{\prime}}{2ABC}\right]f_T-\frac{f}{4}\label{presr}\;, \\
4\pi p_{x} &=& -\left(\frac{C^{\prime}}{4BC}+\frac{A^{\prime}}{4AB}\right)\left(f_T\right)^{\prime}+ \left[\frac{C^{\prime\prime}}{BC}+\frac{A^{\prime\prime}}{AB}-\frac{B^{\prime}C^{\prime}}{2B^2C}+\frac{3A^{\prime}C^{\prime}}{2ABC}-\frac{A^{\prime}B^{\prime}}{2AB^2}-\frac{\left(A^{\prime}\right)^2}{2A^2B}\right]\frac{f_T}{4}-\frac{f}{4}\label{presx}\;,
\end{eqnarray} 
where the equation $x-x$ is the same as $y-y$ ($p_x(r)=p_y(r)$).
\section{Reduction formalism}
In this section by investigating the Riemannian curvature scalar and the scalar torsion we will explain the non equivalence between the results of the Einstein (General relativity (GR) ) and teleparallel gravity (TEGR) at the level of action and also field equations in the presence of a planar source. First, we derive the expression of $R,T$ and the equivalence between the TEGR and GR in the level of action for planar case. The Ricci scalar reads
\begin{eqnarray}
R=-\frac{A''}{AB} +\frac{A'^2}{2A^2B}+\frac{A'B'}{2AB^2}-\frac{A'C'}{ABC}-\frac{2C''}{BC}+\frac{B'C'}{B^2C}+\frac{C'^2}{2BC^2} \label{R}          \,\,.                  
\end{eqnarray}
So, using (\ref{T},\ref{R}) we obtain
\begin{eqnarray}
R+T=-\frac{A''}{AB} +\frac{A'^2}{2A^2B}+\frac{A'B'}{2AB^2}-\frac{2C''}{BC}+\frac{B'C'}{B^2C}+\frac{C'^2}{BC^2}\,\,.
\end{eqnarray}
Then, we have the following action
\begin{eqnarray}
\int{\Big(R\sqrt{-g}+T e\Big)d^4x}\sim\int{dr\Big(\frac{2\sqrt{A}C'^2}{\sqrt{B}C}-C(\frac{A'}{\sqrt{AB}})'-2\sqrt{A}(\frac{C'}{\sqrt{B}})'\Big)}\,\,,
\end{eqnarray}
where we used $e=\sqrt{-g},\int{d^4x}\sim\int{dr}$ over a planar configuration. By integrating by part and assuming that 
\begin{eqnarray}
\lim_{r\rightarrow\infty} \Big(\frac{A'C}{\sqrt{AB}}+\frac{2\sqrt{A}C'}{\sqrt{B}}\Big)=0\,\,,
\end{eqnarray}
we finally obtain
\begin{eqnarray}
\int{\Big(R\sqrt{-g}+T e\Big)d^4x}\sim\int{dr \frac{2C'}{C}\frac{(AC)'}{\sqrt{AB}}}\,\,.
\end{eqnarray}
It means that obviously $R\neq T$ and further we have 
\begin{eqnarray}
R+T\sim\frac{2(AC)'C'}{ABC^2}\neq(\Sigma)'\,\,.
\end{eqnarray}
Therefore, the essential dynamical features of the Einstein case and the TEGR Lagrangian are different and so the general global equivalence between GR and TEGR in plane symmetry is broken. Indeed, this equivalent depends on the symmetry and also the topology and not only the metric (geometry)\cite{hayashi}. Since the plane symmetry can be treated as a special sub-case of a more general axial (precisely it is cylindrical) symmetric objects, the equivalence also will be broken in any other geometry with symmetry (topology) of non spherical shape.

\section{Vacuum solutions with vanishing cosmological constant }
Following \cite{prd} we first classify vacuum solutions with $f(T)=T$. In this case of TEGR, the solution is given by the following general metric
\begin{eqnarray}
ds^2=\Big[\frac{\beta^2}{4\alpha}C(r)+\frac{\alpha}{\sqrt{C(r)}}+\beta\sqrt[4]{C(r)}\Big]dt^2-\Big(\frac{\gamma C'^2}{\sqrt{C(r)}}\Big)dr^2-C(r)(dx^2+dy^2)\,\,.\label{Tsol}
\end{eqnarray}
The form of the metric given in (\ref{Tsol}) makes us ready to investigate the coordinate independent physical properties of the system. The first note is, if we perform a coordinate isomorphism transformation on the coordinate $r\rightarrow r_1=r_1(r)$ then the metric function $C(r)$ preserves the same form as $C(r)\rightarrow C_1(r_1)=C[r(r_1)]$. It can be shown that there exists a unique one to one conformal map (or conformastationary) between the two forms of metric, one with $C(r)$ and another obtain by $C_1(r_1)$, so if we write the metric as $ds^2,d\hat{s}^2$, then the existence of the conformationary map means
\begin{eqnarray}
d\hat{s}^2=e^{\Omega}ds^2\,\,,
\end{eqnarray}
where here the conformal factor must be read as the complete conformal function or particularly as the conformstationary factor. Also, it is easy to show that (\ref{Tsol}) includes also the Minkowskian metric as a special exact (but trivial) solution of the TEGR in plane symmetry. But there is a main difference between these two flat spaces in TEGR and GR: the global properties of the vacuum Minkowskian spacetimes in GR and TEGR are the same. But the local geometry is different because on existence a parameter relates to the topology. This case also occurs in flat spacetimes in the cylindrical symmetric spacetimes, when we investigate the dynamical properties of the cosmic string line element as the exterior solution to the Levi-Civita vacuum solution.
To recover the solution of globally Riemannian zero curvature solution with $R=0$, we take the metric function in the following form:
\begin{eqnarray}
C(r)=\Big[\frac{3}{4\sqrt{\gamma}}(r-r_0)\Big]^{4/3}\,\,.
\end{eqnarray}
For the solution (\ref{Tsol}) the expression of $T$ reduces to the following form:
\begin{eqnarray}
T=\frac{3}{2\gamma}\,{\frac {1 }{ \beta C(r^*) ^{3/4}+2\alpha }}
\end{eqnarray}
In analogue to the GR, the torsion singularities exist in the following points
\begin{eqnarray}
T(r^*)=\infty\Rightarrow C(r^*)  ^{3/4}
 ( \beta C(r^*) ^{3/4}+2\alpha
)=0\,\,\,.
\end{eqnarray}
For a general class of the metric functions $C(r)$, the possible singularities locate at 
\begin{eqnarray}
C(r^*)=0,\ \  C(r^*) =\sqrt[4/3]{\frac{2\alpha}{\beta}}\,\,\,,
\end{eqnarray}
showing the existence of a coordinate singularity. Another particular solutions depend on the specific choices of the metric function $C(r)$. For a more general point of view, the restrictions on the form of metric functions $C(r)$ is just it must be positive at least for $r\geq r_{*}$, monotonic and continuous in a range of the coordinate $r$. The metric of (\ref{Tsol}) represents gravitational configuration of a massive body. The range of the coordinate can be any of these possible intervals :$r\in(r_{*},\infty),\ \ r_{1}\leq r\leq r_{2}$. For example if the metric  denotes a planar analogue of a Schwarzschild-de-Sitter(Anti de-Sitter) metric the last interval of coordinate is refereed.
The proper physical distance is defined by the following integral
\begin{eqnarray}
\hat{r}_{2}=-\int_{0}^{r_2}B^{1/2}dr=\frac{4}{3}\Big[C(0)^{3/4}-C(r_2)^{3/4}\Big],\ \ C'(r)>0\label{rhat}.
\end{eqnarray}
 From this proper physical distance we conclude that the possible singularities locate at the finite distance (physical distance) of the plane. The (\ref{rhat}) explicitly gives us the distance and it depends only on the metric function $C(r)$.
\subsection{Existence of Moller spacetime}
Also, (\ref{Tsol}) produces the Moller solution as a special subclass \cite{moller}. To recover the Moller solution we perform the transformation $C(r)dr=dw$, and it is easy to show that (\ref{Tsol}) recovers the Moller spacetime
\begin{eqnarray}
ds^2=(1+\frac{\beta^2}{4\alpha}w)dt^2-dx^2-dy^2-dw^2\,\,.
\end{eqnarray}
The behaviour of the null geodesics in (\ref{Tsol}) is very interesting. The null photon path gives us
\begin{eqnarray}
t-t_0=\sqrt{\gamma}\int_{r_0}^{r}{\frac{dC}{\sqrt[4]{C}\Big[\frac{\beta^2}{4\alpha}C(r)+\frac{\alpha}{\sqrt{C(r)}}+\beta\sqrt[4]{C(r)}\Big]^{1/2}}}\,\,.
\end{eqnarray} 
For monotonic function $C(r)$ can be integrated to obtain

\begin{eqnarray}
t-t_0&=&\sqrt{\gamma}\sqrt[3]{2}
   \sqrt[3]{\alpha }\frac{4 \left(2 \alpha +\beta  C^{3/4}\right) }{3 \beta ^{4/3} \sqrt[4]{C}
   \sqrt{\frac{\left(2 \alpha +\beta  C^{3/4}\right)^2}{\alpha 
  \sqrt{C}}}}\nonumber\\
  &\times & \Bigg[ \log \left(2 \alpha ^{2/3}-2^{2/3}
   \sqrt[3]{\alpha } \sqrt[3]{\beta } \sqrt[4]{C}+\sqrt[3]{2} \beta
   ^{2/3} \sqrt{C}\right)\nonumber\\
   &-& 2 \log \left(2 \sqrt[3]{\alpha }+2^{2/3}
   \sqrt[3]{\beta } \sqrt[4]{C}\right)+2 \sqrt{3} \tan
   ^{-1}\left(\frac{1-\frac{2^{2/3} \sqrt[3]{\beta }
   \sqrt[4]{C}}{\sqrt[3]{\alpha }}}{\sqrt{3}}\right)+6
   \sqrt[3]{\beta} \sqrt[4]{C}\Bigg]\,.
\end{eqnarray}

Another representation of (\ref{Tsol}) is
\begin{eqnarray}
ds^2=\Big[\frac{\beta^2}{4\alpha}\eta^{4/3}+\alpha\eta^{-2/3}+\beta\eta^{1/3}\Big]dt^2-\tilde{\gamma}^2 d\eta^2-\eta^{4/3}(dx^2+dy^2)\,\,.
\end{eqnarray}
In this coordinate system there is no gauge freedom for metric and the singularity structure of the metric can be investigated in a coordinate independent form. The Torsion singularity can be read from the poles of the following function
\begin{eqnarray}
T=\frac{3}{2\gamma}\,{\frac {1}{\eta
 ( \beta\eta+2\alpha\,\,.
)}}
\end{eqnarray}
The poles read
\begin{equation}
\eta=0,\ \ \eta=-\frac{2\alpha}{\beta}.
\end{equation}
The first singularity is naked singularity and the next one is an horizon singularity. 

\section{Vacuum solutions for  $T=\text{Constant}$}
 In GR, the family of planar constant curvature solutions in vacuum has been previously introduced \cite{horsky,horsky2}. The case of $T=\text{constant}$ is the analogies of such families in generalized teleparallel gravity of us. So, we study this case. If we restrict to $T=T_0$, then the system of Eqs. (\ref{dens}-\ref{presx}) integrated easily for the following set of the metric functions:
 \begin{eqnarray}
 A(r)=C(r)=\frac{1}{B(r)^2}=(1-\sqrt{\frac{T_0}{6}} r)^2\,\,,
 \end{eqnarray}
there exists a large class of $f(T)$ models which can satisfy this metric as the special solution. The viable model of $f(T)$ which satisfies all the field equations is given by the following function
\begin{eqnarray}
f(T)=(\frac{T}{T_0})^{3/4}
\end{eqnarray}
We report this paper before as the special case of the models which posses the Noether symmetry \cite{epjcns}. The viability is easy to be checked because as we showed before, it granted acceleration expansion and also from the cosmographic descriptions. It is interesting that the following metric can produce the de-Sitter solution in planar form as a special representation. In GR, people showed this fact by the use of a Rindler type of coordinates. Such coordinate transformation also here is valid. The solution we obtained for the viable model of $f(T)\sim T^{3/4}$ is singularity free as you can check by direct search of the poles of the torsion scalar (\ref{T}).

\section{Kottler-Wittaker family}
Another solution can be obtained by setting $T=T_0$ in the (\ref{dens}-\ref{presx}) and by carefully investigating the solutions. The following solution also satisfies the field equations in constant torsion case
\begin{eqnarray}
C(r)=1,\ \ A(r)=B(r)^{-1}\,\,.
\end{eqnarray}
This solution corresponds to zero torsion as we see from (\ref{T})
\begin{eqnarray}
T=0\,\,.
\end{eqnarray}
This solution is the torsion-like analogous of the previously solution in GR with $R=0$,  obtained in \cite{wittaker}.
The line element reads
\begin{equation}
ds^2=A(r)dt^2-\frac{dr^2}{A(r)}-dx^2-dy^2\,\,.
\end{equation}
From this metric and also by assuming $T=0$ we obtain
\begin{equation}
A(r)=1-2gr-\lambda r^2\,\,.
\end{equation}
By putting $\lambda=0$ this solution shows the Rindler space time, the spacetime of a moving free falling particle in constant gravitational field, for example the metric a test particle near the Earth. About the algebraic expression of $f(T)$ here, we can propose any function being non singular around $T=0$.

\section{Conclusions}
The existence of planar solutions in a general f(T) gravity has been discussed in this paper. We derived the full system of equations of motion using the tetrad formalism. We showed that at level of action, the GR action and TEGR have different kinds of physics due to a non conservative (not conserved Noether ) charge term. This non local additional term break the equivalence of the linear torsion and Einstein-Hilbert action at level of action. Also, we showed that there exists different family of exact solutions from constant torsion families to the singularity free models. The solutions had different interesting physical meaning. In the case of vacuum constant torsion, the model predicts a $f(T)\sim T^{3/4}$ previously obtained by imposing the Noether symmetry as the point-like symmetry of Lagrangian. Existence of de-Sitter spacetime in planar forms is useful for construct a gauge/gravity duality model in a torsion spacetime.

\vspace{0,25cm}
{\bf Acknowledgement:}  M. E. Rodrigues thanks UFES and UFPA for the hospitality during the development of this work and thanks CNPq for partial financial support. M. J. S. Houndjo thanks CNPq-FAPES for financial support.

\end{document}